# The dwarf galaxy population of the Virgo Cluster
(Review paper presented at the Ringberg Castle workshop on the Virgo cluster)

Noah Brosch
The Wise Observatory and the School of Physics and Astronomy
Beverly and Raymond Sackler Faculty of Exact Sciences
Tel Aviv University, Tel Aviv 69978, Israel

## Abstract

I review the status of knowledge about the dwarf galaxies in the Virgo Cluster (VC) concentrating on work published after 1995. I explain the present ideas about the nature and history of the dwarf elliptical galaxies and emphasize the major increments in understanding achieved in the study of the dwarf irregular population. Special attention is given to the samples of late-type dwarf galaxies in the VC studied at the Wise Observatory (BCDs and dIs), and to a comparison of their properties.

## Introduction.

The VC is the nearest major cluster of galaxies and, as such, it is an ideal place to study various types of galaxies and their interactions. The distance, estimated at 15 to 20 Mpc, implies that fairly faint objects can be reached with modest telescopes and observing time, at least for surface photometry purposes. Additionally, the nearness of the VC to the Milky Way implies that individual studies of galaxies in other spectral bands, e.g., HI or Far Infrared (FIR), can be very detailed. The obvious shortcoming is the large angular extent of the cluster; to cover its core region as well as its Southern Extension one needs to study ~300 square degrees of the sky.

Because of the fortunate accident of our nearness to the VC, the population of dwarf galaxies, objects with $M>-18$ mag, is a particularly attractive subject of study. Dwarf galaxies, and in particular the star-forming irregulars, have been thought to represent a local population that is most similar to the distant, first generation galaxies in the early Universe. The reason is that, observationally, these objects are devoid of large-scale patterns such as spiral arms thus the traditional star-formation triggers are probably not active. In addition, many dwarf irregulars have been found to have low metal abundances with the record being 1/52 solar for IZw18.

The landmark work dealing with the general population of the VC is undoubtedly the Las Campanas survey of galaxies (VCC catalog: Binggeli et al. 1985). The wide scale of the photographic plates used at the 2.5-m telescope allowed fairly confident morphological classification of more than 2000 objects. Among the dwarf galaxy (DG) population, Binggeli et al. identified approximately 1000 such objects that are considered cluster members; 90% of these are dwarf ellipticals (dE) and the remaining 10% are classified as either dwarf irregulars (dI) or blue compact dwarfs (BCD). The VC membership assignment by Binggeli et al. was based on morphology; this was checked later using redshifts (e.g., Drinkwater et al. 1996) and was found to be mostly correct.

A second landmark contribution to the study of the VC was by Hoffman et al. (1987, 1989), in the Arecibo survey of all late-type galaxies in the VCC catalog. Although at times missing flux, because of the single beam observations, the HI survey is important because it (a) allowed the elimination of some late-type dwarfs from the list of VC members, by finding that these were background objects, and (b) yielding the total HI content and an indication about the dynamics of the HI from the line profile. Parenthetically, in most of the cases of dI and BCD galaxies the profiles measured at Arecibo were far from typical of disks; no two-horned profiles were observed for bona-fide members of these classes arguing against a disk explanation for these galaxies.

The third landmark contribution allowed the evaluation of the influence of hot intra-cluster gas (ICM) to the galaxy evolution. The major contribution came from the ROSAT survey of the VC (Böhringer et al. 1994) with some contribution from the EUVE observations (Berghöfer et al. 2000) where lower temperature gas was targeted. The latest EUV maps indicate that this kind of emission is relegated to the immediate neighborhood of M87 and requires the presence there of relativistic electrons. In the context f a hot ICM, I want to mention the ASCA "mystery hotspot" showed at this meeting by Hans Böhringer, where a region ~4 square degrees wide, somewhat to the South-West of M49, shows much higher temperatures than any region in the VC: 4.5 keV vs. a typical cluster temperature of 2.5 keV (see also Shibata et al. 2001). This is suggested by Kikuchi et al. (2000) to be the result of infall of the M49 group toward M87 with ~1000 km/sec, causing a shock in the ISM.

<u>Dwarf and large galaxies.</u>

In principle, galaxies can be classified in a two-dimensional diagram of total luminosity vs. surface brightness, where they concentrate in two different and ~orthogonal branches. The spiral branch contains also the irregular galaxies, the dwarfs of various flavors, and the dwarf spheroidals. It is well separated from the elliptical branch, which contains also the cDs at one end and the low luminosity ellipticals at the other. Weird and rare galaxies, such as the low surface brightness objects and the "Malin I" types, as well as the BCDs, deviate by various amounts from these, apparently well-defined, sequences.

Problems appear when considering the proposed population of dwarf spirals, a class of ~6 kpc and low luminosity objects, or galaxies that appear elliptical but image processing reveals spiral arms. Are these the dwarf lenticulars (dS0) introduced by Sandage & Binggeli (1984)? Among the dEs there are at least two kinds of flavors: there are "disky dEs" and "bulgy dEs" according to the shape of their outer isophotes, and there are nucleated dEs (dE,N) as well as pure dEs. A revisit of the photographic material analyzed by Binggeli et al. (1985) yielded the interesting statistic that while 10% of the dEs at absolute magnitude –12 are nucleated, this fraction grows to almost 100% for dEs brighter than –16 (Binggeli et al. 2000).

Oh & Lin (2000) simulated the fate of globular clusters (GCs) around dwarf ellipticals and showed that they will eventually sink to the centers of these dEs, transforming them into the nucleated version. This process apparently takes a few Gyrs and happens mostly in the inner regions of a cluster. The sinking of GCs may explain also the off-center position found by Binggeli et al (2000) for the nuclei of dEs but contradicts (a) the nuclear magnitudes measured by Durret (1997), which at −10 mag are too bright to be GCs, as well as (b) the preliminary results from the HST snapshot program to evaluate the frequency of GCs near Es (Miller et al. 1996). The latter found that dE,N galaxies tend to have more GCs than their non-nucleated cousins, rather the opposite of what one could expect.

The contribution of DGs to the total population is normally evaluated through the luminosity function. Here the literature sees significant confusion regarding the VC, primarily at the faint end of the distribution. While Caldwell & Armandroff (2000) claim a slope of −1.2 for the faint objects, Phillipps et al. (1998) estimate a faint-end slope of −2. At this meeting, Sabina Sabatini claimed an even steeper slope of −2.2 from a blind survey of some ten square degrees extending from M87 westward, significantly enhancing the number of low luminosity and low surface brightness galaxies in the cluster. Tully (private communication) supports a lower faint-end slope for the Virgo luminosity function and indicated that a flat low-luminosity end is seen also for the spiral-rich Ursa Major cluster at a distance from us that is similar to that of the VC. The differences in luminosity functions result from the difficulty of accounting properly for the contribution of background galaxies, in absence of redshift information, primarily for the low surface brightness galaxies.

Note also that, at present, there is no claim of any correlation between the proposed low surface brightness galaxies discovered by the Cardiff group in the direction of the Virgo cluster and any of the intracluster planetary nebulae (PN) candidates (e.g., Arnaboldi et al. 2002), although one would expect that with ∼300 PNs and ∼5 LSB galaxies per square degree, some PNs should be visible in these objects allowing an unbiased distance determination.

The dynamical properties of the different galaxy populations were recently studied by Conselice et al. (2001) for the VC core, a region six degrees in diameter around M87. This shows that the dEs form a population different from that of the other galaxies, or that they are a mixture of two different populations; one at the typical cluster redshift and another at a higher redshift band, about 1000 km/sec away from the cluster redshift. From this, Conselice et al. claim that the dEs were accreted by the VC not as dEs, but as a different type of progenitor galaxies (perhaps spirals, cf. Conselice et al. 2000). This was also proposed by Mao & Mo (1998) on theoretical grounds.

This general picture of DGs in the VC raises (at least) the following questions: (a) what is the relative importance of nature vs. nurture in DGs, (b) are the VC DGs the building blocks of galaxies or are they parts of large galaxies torn off during violent cluster-related processes, (c) what happens to DGs as they accrete onto the VC, (d) what is the fate of the debris from disrupted DGs, and (e) what process triggers star formation (SF) in DGs. Some of these questions may have answers in the present workshop.

<u>The Wise Observatory project.</u>

This project represents an investment of ∼15 person-years in observations and follow-up interpretation. It studied well-selected samples of late-type dwarf galaxies in the VC, consisting of a representative selection of BCDs and a complete sample of dIs. The BCDs were selected to have HI detections at the redshift range of the VC and the dIs had to belong to the ImIV or ImV flavors (low surface brightness=LSB), and have similar restrictions in the total HI content. The studies were done using CCD imaging in (U)BVRI and H$\alpha$ (at the galaxy rest frame) and yielded information about the morphology, the total and localized SF, the history of SF, etc. The observational results were described in the PhD theses of Almoznino and of Heller, as well as in the follow-up publications.

A brief summary of the findings from the WO project follows. The luminosity distribution for the low surface brightness galaxies follows a Sérsic profile with exponents ranging from 1 to 2 (median 1.74±0.86, Heller et al. 2001). In particular, no deVaucouleurs profiles could be fitted to any galaxy. In most cases exponential, or truncated exponential profiles seem to be the law. This is similar to what Ryden et al. (1999) found for a sample of dEs and is reminiscent of the finding of Gavazzi et al. (2001) that dEs and late-type dwarfs have similar structural properties when studied in the H-band (1.65 µm). We studied the spatial distribution of light in the broad-band images and found that while the HSB dwarfs show a color gradient from the center to the periphery (the center being bluer), this is not detected in the LSB galaxies (Heller et al. 2001).

The distribution of H$\alpha$ emission shows that in LSBDGs the SF takes place mostly at the edge of a galaxy, and mostly to one side (lopsidedness property: Heller et al. 2000). This is in contrast with the HSB sample (BCDs), where the SF is mostly central, or is relegated to a small number of HII regions. This fits the behavior of the broad-band color gradient mentioned above.

The imaging photometry in H$\alpha$ (Heller et al. 1999) showed also that the total SF rate (SFR) is modest in both types of galaxies: $6.5 \times 10^{-2} M_{Sun}/yr$ for BCDs and $7 \times 10^{-3} M_{Sun}/yr$ for the dIs. However, compared with large (i.e., non-dwarf) spirals the specific SFR, that is, the SFR per unit of blue luminosity is rather high in both cases and is commensurate with the SFR in the Milky Way. The SFRs distributions for the two populations overlap and there are dIs with higher SFRs than some BCDs. The underlying brightness of the red continuum, under the detected HII regions, correlates with the H$\alpha$ flux from the specific HII region. This correlation is shown in Figure 1 and indicates that the triggering of SF is probably by some local property, such as the local mass of old stars, or a local, long-lasting concentration of dark matter.

In addition, note that we could not identify any VC-related property that correlated with the SFR. Neither the Virgocentric distance, nor the radial velocity of a DG, seems to have an influence on the SFRs. We could not identify immediate neighbors with whom a DG could have interacted, thus we can rule out prompt tidal interactions as triggers of SF. This is similar to the findings reported at this workshop by Yun. The asymmetry and concentration properties of the H$\alpha$ emission can be reproduced by a random distribution of HII regions (Heller et al. 2000); this indicates that probably a variant of the Gerola et al. (1980) stochastic star formation scenario may be working in these LSB dIs. For HSBs, there is a strong tendency to show a single HII region, or a small number of HII regions.

We compared our broad-band and Hα photometry with predictions from different evolutionary stellar populations in order to derive plausible SF histories for these galaxies. We used models from Bruzual & Charlot (1993, 1995), and from Mas-Hesse & Kunth (1991). The latter is crucial in understanding galaxies with current star formation, because the models of Bruzual & Charlot do not deal with light contributions from hot gas. An example of a diagnostic diagram, using Hα, B, and V, is shown in Figure 2 (from Heller et al 2002).

The figure shows that simple, single-population models cannot satisfy the observations. We found that combinations of two widely different populations, in various proportions, can fit all the diagnostic diagrams. The two populations must have been formed in short bursts, at least the recent one must have low metallicity, and the two bursts must be spaced by at least a few hundreds of Myrs up to a few Gyrs.

Spectroscopic studies and the question of gas stripping.

In principle, much information about modes of star formation can be obtained from spectroscopy of the HII regions, particularly by evaluating their metallicities. The recent contribution by Lee (2000) stands out, because he compared the metallicities of HII regions in 12 VCdIs with those of a comparison sample of dIs in the field.

Lee (2000) derived a method of predicting the total HI content of a galaxy from its oxygen abundance. This allowed him to establish the HI deficiency of dIs in the VC, which he found to correlate with the local X-ray flux. According to Lee, this is a supporting finding to the assertion that HI is stripped from galaxies, and in this case dIs, by ram interaction with the hot intracluster medium. Lee also identified an isolated HI cloud near VCC 1249 containing a low luminosity dI; he proposes this to be the result of a tidal interaction between VCC 1249 and M49, which loosened the ISM in the dwarf galaxy, followed by ram pressure stripping, the formation of a separate entity HI cloud, and subsequent star formation (see also Lee et al. 2000).

The issue of ram pressure stripping was discussed extensively at this meeting, mainly in the context of spiral galaxies. The case of NGC 4338 may be another example of stripping for the case of a dwarf galaxy (Yoshida et al. 2002). Also, the strange features shown here by Manfred Stickel regarding M86 and its dwarf neighbor VCC 882 may also point to stripping, although some kind of tidal interaction cannot be ruled out. The external HI contours, the dust feature located between the two HI peaks (Elmegreen et al. 2000), and perhaps VCC 882 itself, may all originate from an interaction of NGC 4402 with M86. Obviously, detailed modeling that could account for all the ingredients (stars, dust, and HI gas) is required for this object.

Another piece of evidence supporting the claim of HI gas removal from cluster dwarf galaxies can be found in the recent survey of HI in compact galaxies located in voids (Pustilnik et al. 2002). It was shown there that the dependence of the total HI content on the luminosity of a compact galaxy depends on the local galaxy density. This is emphasized in the two panels of Figure 3. The top panel shows the relation for the Virgo BCDs in the Almoznino & Brosch (1998) sample. The bottom panel shows it for a combined sample of compact galaxies selected to be at large distances from their nearest neighbors, and considered to represent a void population of compact galaxies.

The comparison is done through the β parameter, defined from the relation $M(HI)/L_B \propto L_B^{\beta}$. Fits to the distribution of points show that β=0.07 for Virgo BCDs while β=-0.40 for void BCDs. The local neighborhood density of galaxies seems to influence the ability of a galaxy to retain its HI not only in the dense cores of clusters, but also in the typical regions of the Local Universe.

### ISM in DGs and other relevant questions.

The question of the interstellar matter in dwarf galaxies was emphasized recently by the very interesting series of papers resulting from the ISOPHOT survey described here by Tuffs. The intriguing result regarding extremely cold dust, with T<10K, located in or around BCDs (Popescu et al. 2002) raises an number of questions, such as (a) what is the nature of the detected dust, i.e., is this really Galactic dust, in such large quantities as claimed and located in metal-poor galaxies, and (b) what are its origins, i.e., is this dust internally produced during a previous episode of SF or is it accreted from within the cluster. It is not clear whether any of this dust may be connected with extended HI distributions (e.g., Hoffman et al. 1996). Note also that the other ISO contribution on the subject of dust in galaxies (Boselli et al. 1998) showed the dearth of the very small grains or grain inclusions that are responsible for the mid-IR emission at 6.75 and 15 μm from low-mass dwarfs. There, this was explained as due to the destruction of this dust component by the UV photons, but it is possible that the carriers of these mid-IR features do not form at all in metal-poor DGs. In this context, note that Gondhalekar et al. (1998) did not detect CO emission from some members of a sample of compact and star-bursting galaxies. The CO emission was absent in the low-luminosity members of the sample, essentially those that would be classified as dwarfs. This also may point out a link with the low-metallicity property of these objects.

A final remark about dwarf galaxies concerns the ability of modern large-scale structure simulations to reproduce the dwarf galaxy population. Realizations of the Local Universe, e.g., Klypin et al. (2001), apparently reproduce well the Local Supercluster, Virgo, Coma, and Perseus-Pisces, both in terms of the mass distribution and in terms of the velocity fields (large scale flows). These simulations can now have very fine spatial resolution (~20/h kpc) and good mass resolution that can go as low as 40 km/sec. This is the dispersion velocity of small-mass halos, which presumably can simulated dwarf galaxies. The theoretical discipline could, therefore, simulate the local population of dwarf galaxies in and around the VC and throw some light on the question of Virgo infall and the formation of the dE and dI populations.

Regarding studies of galaxies in the Virgo cluster, and specifically the determination of the influence of the local, galactic ISM on X-ray measurements, I want to emphasize a caveat regarding the proper accounting for foreground material. We have shown (Bosch et al. 1999) that FIR observations in the direction of the center of the Virgo cluster indicate the presence of significant quantities of dust. Some of this dust does not seem to cause reddening of background galaxies, indicating probably the presence of large dust grains that produce gray extinction. If this is the case, then usual methods for detecting such dust and correcting for its influence (e.g., as done for the Cepheids in the HST Key Project) will underestimate the extinction and will produce wrong distance estimates. This is true also in the case of dwarf galaxies, and is difficult to properly account for this effect.

Future observations.

It is possible to compile a list of observational wishes by combining whatever is known on DGs in general and what could be expected to be found. Such observations would include an extension of the FIR survey to all the late-type DGs in the VC, which would be deeper and more comprehensive than the Tuffs et al (2002) survey. In addition, we would desire a thorough survey to the DGs in the near-IR; this would allow the understanding of the old stellar population, better than what the R or I bands are able to constrain. More HI maps, at least for the HI-rich objects, would allow a better understanding of the internal ISM dynamics; this should be combined with H$\alpha$ mapping with an imaging Fabry-Perot interferometer to derive the relation between the neutral and the ionized gas dynamics. More slit spectroscopy should be obtained now that detailed information about the location of HII regions is available from the H$\alpha$ images. This would allow the tracing of the metal enrichment within the disk of the galaxy and should be combined with the internal dynamics. In the same line, mapping of the CO and other molecules would also be useful when determining the history of metal enrichment. Finally, all these observational results are only useful if a comparison sample of non-cluster galaxies is available. The definition of such a sample, that would be acceptable by the community, should take a high precedence.

Conclusions.

The dwarf galaxy population of the Virgo cluster is dominated by dwarf ellipticals. These may have been accreted by the cluster as different types of galaxies, and may have been transformed into dEs through some interaction with the ICM or with other cluster galaxies. A significant fraction (~10%) of the DGs belongs to the dI class and shows, in many cases, signs of current star formation.

Various authors proposed that at least the dEs result from an accreted population of progenitor galaxies; presently the nature of these predecessor objects is not clear, nor is there a visible reservoir of such galaxies in a pre-accretion phase. We therefore must await a better explanation for the dEs, or another idea about their origin.

Our extensive studies of late-type dwarf galaxies in the VC show that the proposed scenario relating BCDs to LSB dwarfs (lazy galaxies that sometimes flare-up) is not tenable. There is no way by which an Im IV-V object can become a BCD, and vice-versa. However, there are cluster-related mechanisms by which a galaxy may lose part of its ISM; these are harassment, tidal interactions, and ram-pressure sweeping, and may be more relevant to dwarf galaxies because of their shallower gravitational potential.

Acknowledgements

I am grateful to Ana Beatriz Heller and Elchanan Almoznino for major contributions to the study of the Virgo cluster dwarf irregular populations. At various stages of this project G. Lyle Hoffman helped with data and Ed Salpeter discussed some of the results. Data and remarks by John Salzer and Liese van Zee are greatly appreciated. Part of the work reported here was done while I was a sabbatical visitor at the Space Telescope Science Institute; I am grateful to Bob Williams, its past Director for facilitating this extended stay. Research at the Wise Observatory is supported by grants from the Israel Science Foundation and from the Austrian Friends of Tel Aviv University.

Figure captions:

Figure 1: Correlation between the underlying red continuum flux from HII regions and their H$\alpha$ fluxes. HII regions in BCDs are plotted with open circles and those in dIs are plotted as stars. Lines of different H$\alpha$ equivalent widths are also plotted.

Figure 2: Diagnostic diagram from Heller et al. (2002). Individual galaxies are plotted as filled triangles and are labeled with their VCC number. The models are plotted as lines and represent a single stellar population (ssp) at different ages and burst populations at different ages and with various metallicities. Lines representing a mix of the two populations connect the ssp model and one of the recent burst models. The fraction of light contributed by the very young population is parameterized as $k$.

Figure 3: The HI mass to blue light ratio vs. the absolute B magnitude for a sample of extremely isolated (void) compact galaxies (lower panel) and a sample of BCDs in Virgo. The solid line in each panel is the best-fit relation for the specific sample described by the panel. The dotted lines are best fits for other samples considered by Pustilnik et al. (2002), two of which (not shown here) are of compact galaxies in the Local Supercluster and in the General Field.